# Laboratory Investigations of the Dynamics of Anomalous Entrainment in Cumulus-cloud flows

Sourabh S. Diwan[1], Roddam Narasimha[1], G.S. Bhat[2] and K.R. Sreenivas[1]

[1] Jawaharlal Nehru Centre for Advanced Scientific Research, Bangalore

[2] Indian Institute of Science, Bangalore

**Abstract**

Anomalous entrainment in cumulus clouds has been a topic of investigation over many decades in the past. Its importance stems from the fact that entrainment rate is one of the major inputs to several cumulus-parameterization schemes. Recently Narasimha *et al.* (PNAS; 2011) have successfully simulated the large-scale dynamics of cumulus-cloud flows in the laboratory and provided a mechanistic explanation for the observed cumulus-entrainment anomalies. They showed a favourable comparison of a dilution-related quantity (called 'purity') between the laboratory measurements and cloud-resolving-model computations, and discussed the important problem of homogeneous vs. inhomogeneous mixing in cumulus clouds. The main purpose of the present report is to provide additional supporting information and a more detailed account of the entrainment-related issues not included in the PNAS paper due to space constraints. We believe this report, in conjunction with the paper, will present to the reader a comprehensive and more-or-less complete documentation on the issues mentioned above.

## 1. Introduction

Cumulus clouds, whose science involves a complex interplay among dynamics, thermodynamics, microphysics, radiation etc., represent the largest source of uncertainty in weather and climate modelling. Understanding cloud physics and dynamics is therefore a topic of intense current investigation, using chiefly field measurements and large-eddy simulations (see e.g. Blyth *et al.* 1988, Gerber *et al.* 2008, Siebesma and Cuijpers 1995, Romps and Kuang 2010). The laboratory experiments have mainly focussed on cloud microphysical studies and very few studies have attempted to simulate the *dynamics* of cumulus clouds in the laboratory (Stratmann *et al.* 2009). A new approach in this direction has recently been reported by Narasimha *et al.* (2011), who showed how the macro-scale evolution of cumulus clouds can be simulated in a laboratory apparatus designed by Bhat and Narasimha (1996). They proposed the transient diabatic plume as an appropriate fluid-dynamical model for studying cumulus flow dynamics and explained, for the first time, the 'anomalous' behaviour of entrainment in cumulus clouds. Narasimha *et al.* (2011; to be referred to as N+ in the rest of this report) presented the striking variation of the entrainment coefficient with height in the relevant earlier measurements made on steady diabatic jets and plumes. This result showed that the 'self-preservation' theory of entrainment in cumulus-type flows was untenable. They further showed that the dilution rates (measured using a term called 'purity') found in the laboratory diabatic plumes compare favourably with those obtained in the numerical simulation of steady deep clouds performed by Romps and Kuang (2010). Moreover, N+ pointed out how their laboratory simulations can have implications in understanding the issue of homogeneous vs. inhomogeneous mixing in cumulus clouds, which also is a subject of great current interest.

The present report is meant to provide additional information, apart from the supporting material accompanying the main text in N+, with regard to the issues mentioned above. Section 2 includes a detailed account of a critical re-analysis of the experimental data on steady diabatic jets and plumes, including the assumptions made and the data smoothing performed during the course of the analysis. In section 3 we present the reasoning employed in arriving at the laboratory analogue of purity (which we shall call 'diabatic purity') computed by Romps and Kuang (2010). Section 4 deals with the estimation of the turbulent mixing time scales necessary for deciding the nature of mixing in clouds. In this section, we provide some additional arguments and information to support the proposal made in N+ that mixing in cumulus clouds tends to become more homogeneous with increase in height above the cloud base. A summary is given in section 5.

## 2. A critical re-analysis of available experimental data for calculation of entrainment coefficients in steady diabatic jets and plumes

In this section, we present the methodology we have used in the calculation of entrainment coefficients in steady-state round jets and plumes subjected to off-source heat addition (reported in N+). The entrainment coefficient is defined as (Turner 1973)

$$\alpha_E = \frac{dm/dz}{2\pi\rho b U_c}, \quad (2.1)$$

where *m* is time-mean mass-flow rate integrated in radial direction, *z* is vertical coordinate, *b* is velocity width (it can either be $b_u$, where $U(b_u) = U_c/2$ or $b_{ue}$, where $U(b_{ue}) = U_c/e$), *U* and $U_c$ are mean local and centreline velocities respectively, and $\rho$ is fluid density. Note that the term velocity width used here refers to the radial width (or half the diametral width). Herein, we include five data sets obtained in three different laboratories, viz., Indian Institute of Science Bangalore (IISc), Delaware University (DU) and Florida State University (FSU) (all using a setup very similar to that developed by Bhat and Narasimha (1996) to calculate $\alpha_E$ values as a function of *z*. These data sets have been reported in the papers listed below.

1. Bhat & Narasimha (1996; **IISc**)
2. Venkatakrishnan (1997, *et al.* 1999; **IISc**)
3. Agrawal & Prasad (2004; **DU**)
4. Venkatakrishnan, Elavarasan, Bhat, Krothapalli & Lourenco (2003; **FSU**)

An accurate calculation of $\alpha_E$ requires accurate measurements of mass-flow rate and velocity width. However, since the setup used in the above investigations does not allow run times more than 15 to 20 minutes (and also due to some other reasons to be mentioned below), there is some scatter and uncertainty in the data reported in these studies. As a result, calculation of $\alpha_E$ poses problems and may sometimes show unrealistically large fluctuations as the derivative operator on mass-flow rate (equation 2.1) amplifies small variations. This can result in a non-smooth behaviour in the axial variation of $\alpha_E$, which could be due to one or more of the following reasons (see also N+); (i) insufficient averaging times, (ii) difficulties encountered in measuring small velocities close to the jet/plume edges, and (iii) the presence of heating grids in the measurement zone causing unwanted reflections. Therefore, to get reasonably good estimates of entrainment coefficients, some smoothing and fairing of the raw data (as reported in the above studies) was found necessary. In the following, we present a detailed discussion regarding the step-by-step procedures used (including the reasoning behind each step) to arrive at the estimates of the $\alpha_E$ values in each of these studies. (Note that in these studies *m* (mass/time) has been termed as 'mass flux'. In this section, we call it 'mass-flow rate' to avoid confusion with the definition of mass flux, i.e. mass/area/time, which is commonly used in the cloud-physics literature.)

**2.1 Bhat and Narasimha (1996; BN)**

BN carried out flow visualization and laser Doppler velocimetry (LDV) measurements on steady diabatic jets, which were subjected to off-source heating in a heat-injection zone (HIZ) in $z_b < z < z_t$; here $z_b$ is the beginning of the heating zone and $z_t$ is the end of the heating zone. The non-dimensional parameter BN proposed to characterize the flow, called by them the heat-release number (which can also be interpreted as a bulk Richardson number, BN1996) *G,* to be called $\bar{G}$ in this report to distinguish it from the heat-release number referred to the base of the heating zone, $z_b$, to be defined later in this section (see equation 2.3). The heat-release number is expressed as (equation 4 in BN),

$$\bar{G} = \frac{\beta_T g}{\rho C_p} \frac{z_b^2}{d^3} \frac{Q}{U_o^3} \qquad (2.2)$$

Here subscript *b* indicates the base of the HIZ, $\beta_T$ is the coefficient of volumetric expansion, *g* is the acceleration due to gravity, $\rho$ is the density, $C_p$ is the specific heat at constant pressure, *Q* is the total heat added in the HIZ, *d* is the orifice diameter and $U_o$ is the orifice exit velocity. Ideally, for the calculation of $\alpha_E$, centreline velocity and width data (for a Gaussian mean-velocity profile) should be used at the same value of $\bar{G}$. However, since both the data at the same $\bar{G}$ are not reported in BN, we have chosen centreline-velocity values at $\bar{G} = 4.2$ and velocity-width values at $\bar{G} = 4.4$. We expect that the small difference in the numerical value of $\overline{G}$ will not affect the trend in $\alpha_E$ to leading order, as is borne out by the following exercise.

### 2.1-(a) Centreline Velocity ($U_c$)

The centreline-velocity values were extracted from figure 9 (a) in BN for $\overline{G} = 4.2$ and are reproduced in figure 1 below. (For extracting data from figures, a free software called WINDIG was used throughout this exercise.) The square symbols show the raw data for the heated jet. It was observed that below the HIZ, i.e. $(z/d) < 133$, there was some scatter in the data (see figure 1; note that the scatter is more highlighted in the $1/U_c$ plot than in the $U_c$ plot, not shown here). To smooth out the trend, we make use of the fact that the jet prior to the HIZ follows classical similarity theory i.e. $U_c/U_o = B_u d/(z - z_o)$. Here $d$ and $U_o$ are orifice exit diameter and velocity respectively, $B_u$ is a proportionality constant and $z_o$ is the virtual origin of the jet (or the plume). This relation is plotted in figure 1 as a solid line with $B_u = 5.7$ and $z_o = -4$ (see table 1 in BN). These values of $B_u$ and $z_o$ are valid up to $z/d < 100$ as noted in BN; we have, however, used them till $z/d \approx 120$ on the assumption that the departure from this relation for $100 < z/d < 120$ will be small. The solid line has a slight offset from the trend indicated by the data points below the heating zone (square symbols). A dashed line is therefore drawn parallel to the solid line removing the offset; it passes through the data points at $z/d \approx 71$ and $z/d \approx 95$, while the data points at $z/d \approx 40$ and $z/d \approx 120$ lie slightly away from this line. To obtain a smoother variation of the centreline velocity, the values of $U_o / U_c$ at $z/d \approx 40$ and $z/d \approx 120$ are taken to lie on the dashed line, leaving the rest of the data unaltered. The circles in figure 1 indicate the refined data.

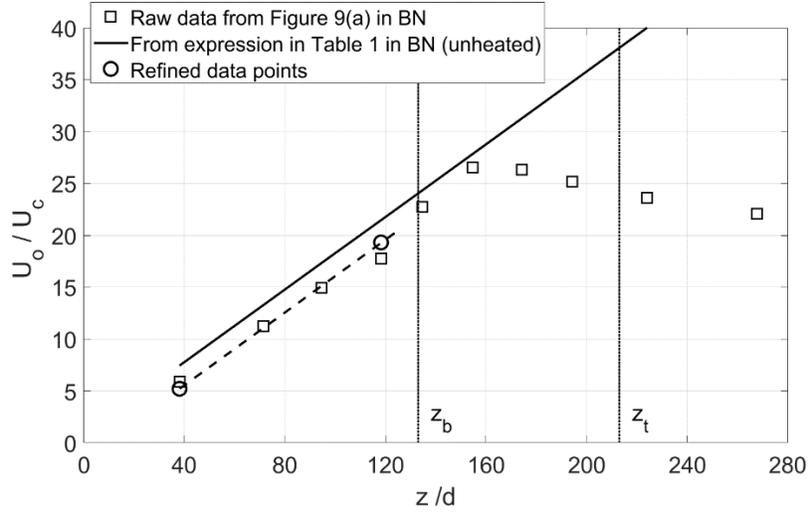

Figure 1 Centreline velocity values extracted from BN. The solid line is based on $U_c/U_o = B_u d/(z - z_o)$ with $B_u = 5.7$ and $z_o = -4$. $z_b$ and $z_t$ indicate the beginning and end of the HIZ; $\overline{G} = 4.2$.

## 2.1-(b) Velocity Width ($b_{ue}$)

BN have reported both scalar ($b_{se}$) and velocity ($b_{ue}$) widths, where the mean pixel intensity (in the visualized images) and axial velocity reach 1/e of their respective centreline values. As already noted above, since measurement of low velocities away from the axis presents difficulties, only a few data points of velocity widths (from direct measurements) are available. For example, figure 11 in BN gives velocity widths at $\overline{G} = 4.4$ at three axial locations; these are reproduced in figure 2 here as red squares.

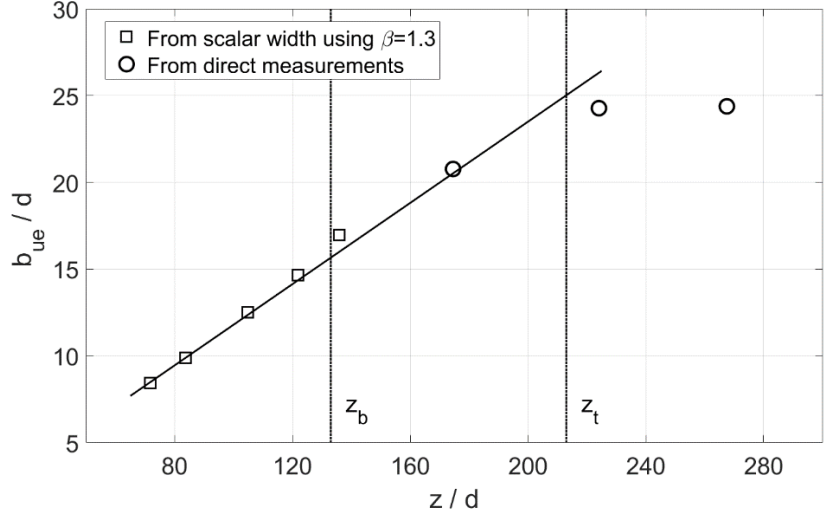

Figure 2. Velocity-width values extracted from BN; $z_b$ and $z_t$ indicate the beginning and end of the HIZ. Note that for the squares, $\bar{G} = 4.7$ and for the circles, $\bar{G} = 4.4$.

Unfortunately, direct measurements of $b$ are necessary for calculation of mass-flow rate at these locations, but are not available near the beginning of and below the HIZ. BN, however, have given scalar-width data in great detail (figure 6 in BN) covering the entire region of interest. We, therefore, have inferred velocity widths from scalar widths ($b_s$) for $z$ locations below and close to the beginning of the HIZ. For this purpose, we chose $\bar{G} = 4.7$, which is close to $\bar{G} = 4.4$ for which direct measurements of $b_{ue}$ are available (shown as circles in figure 2). This is justified since for $z < z_b$ the jet is seen to follow self-similarity laws, and the data points for different values of $\bar{G}$ collapse well on top of each other (figure 6 in BN). BN have given a value of $\beta = b_{se}/b_{ue} = 1.3$ for the unheated jet; the values of $b_{ue}$ obtained from this relation are shown in figure 2 as square symbols. Note that we have not converted $b_{se}$ to $b_{ue}$ for $z > z_b$ (except for the data point corresponding to the last square symbol which is very close to $z_b$), since the dependence of $\beta$ on the heat added in the HIZ is not yet clearly known. The resulting composite variation of $b_{ue}$ with $z$ is seen in figure 2. It is consistent with the general trend that the width first increases beyond the value corresponding to the unheated case (shown by the solid line) and later on drops below it (see also the discussion in section 2.3-(b)). This justifies the present exercise of putting together velocity-width data partly obtained from direct velocity measurement and partly inferred from the scalar-concentration measurement.

**2.1-(c) Mass-flow rate (*m*)**

BN found that their measured (and scaled) axial-velocity profiles slightly beyond the HIZ and dye-concentration profiles inside as well as above the HIZ followed the Gaussian distribution reasonably well for most of the radial extent. Since the mass-flow-rates obtained by integrating the velocity profile in the radial direction are not available, we calculate mass-flow rate at each *z* location as $m = \rho \pi b_{ue}^2 U_c$, assuming the velocity profiles to be Gaussian in shape. For this purpose, values of $b_{ue}$ and $U_c$ were interpolated using 'Shape Preserving Spline' interpolation scheme in MATLAB, and the resulting mass-flow-rate (per unit density; $m/\rho$) variation is shown in figure 3 (a).

The solid line in figure 3(a) shows the mass-flow-rate values for an unheated jet. It is clear from the figure that the effect of heating is to increase the mass-flow rate from the corresponding unheated value inside and beyond the HIZ. This is consistent with the mass-flow-rate variation shown in Narasimha and Bhat (2008) which was obtained by using a variable-β model to convert scalar width into velocity width for $z > z_b$ (see figure 3(b)). Thus, the present exercise supports the contention of Narasimha and Bhat (2008) that the mass-flow-rate values for the heated jet in the experiments of BN are indeed higher than those for the unheated jet (for $z > z_b$) and *not* lower as concluded in Agrawal and Prasad (2004).

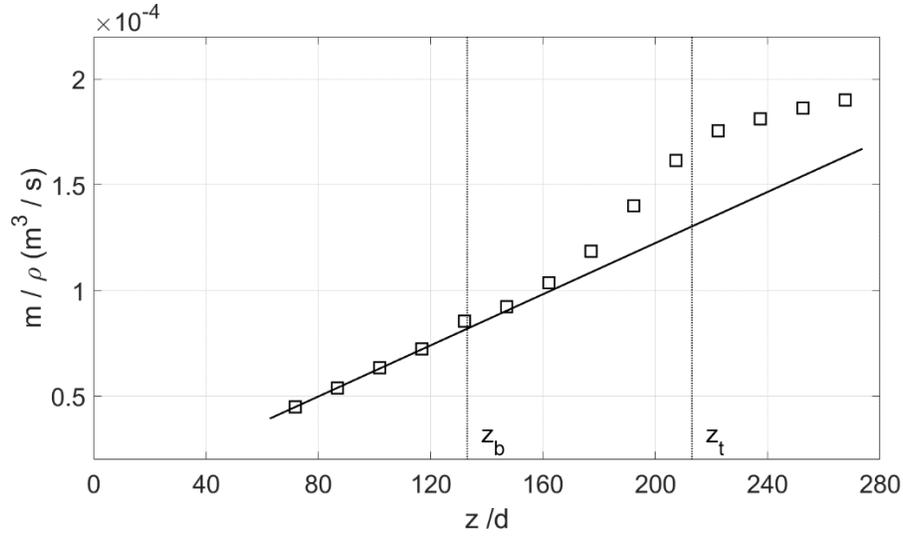

(a)

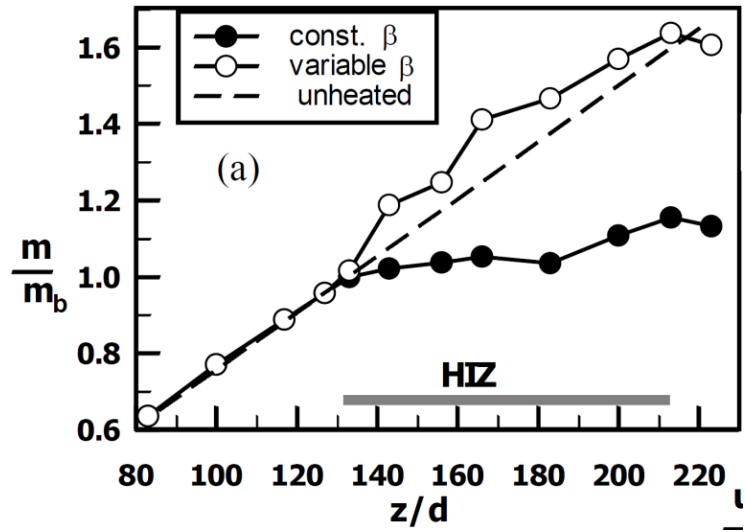

(b)

Figure 3 (a) Variation of volume-flow rate (mass-flow rate per unit density) with $z$ corresponding to the data in figures 1 and 2. (b) Mass-flow rate obtained using variable $\beta$ model; reproduced from Narasimha and Bhat (2008).

## 2.1-(d) Entrainment Coefficient ($\alpha_E$)

The entrainment coefficient obtained from equation (2.1) is plotted in figure 4 below.

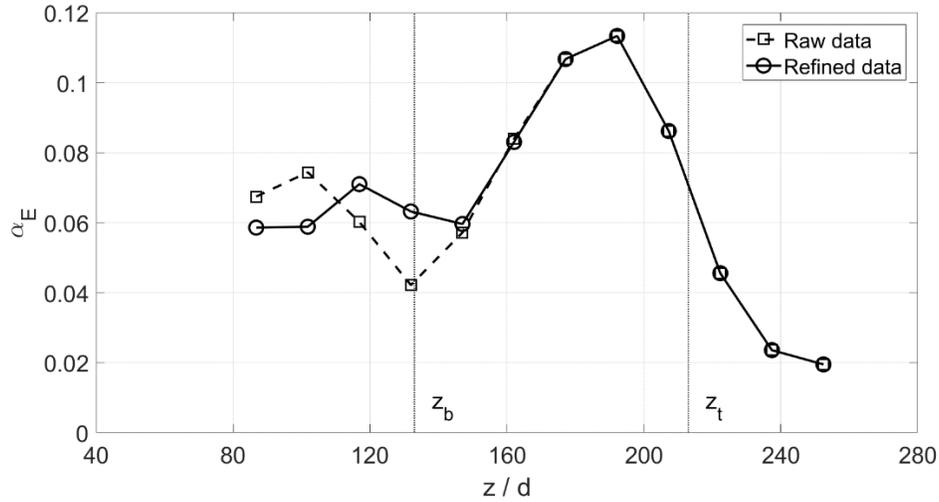

Figure 4. Entrainment coefficient as a function of $z$ for the data in figures 1, 2 and 3 (for both raw and refined values) for the measurements reported in BN.

It is seen that the refined values of $\alpha_E$ are different from the corresponding raw values for $z \leq z_b$; for $z > z_b$, they are virtually the same (as expected). (Note that the term 'Refined values' has been used throughout this section to denote the values obtained by employing the smoothing exercise. For example, in the present case this corresponds to the centreline-velocity variation with its values modified at two locations, $z/d \approx 40$ and $z/d \approx 120$; see figure 1.) In the unheated region, the first two (refined) values of $\alpha_E$ are about 0.058 which are close to the values reported in the literature for a classical jet, e.g. 0.056 as quoted by Turner (1973) and 0.057 as calculated by Hussain *et al.* (1994). (Note that $\alpha_E = 0.054$ given in Turner (1986) is slightly lower than these values.) Thus the $\alpha_E$ value obtained for the data of BN is consistent with the results of the previous studies.

## 2.2 Venkatakrishnan (1997 – PhD thesis; $V_T$)

Venkatakrishnan (**$V_T$**) carried out flow visualisation and Laser Doppler velocimetry (LDV) measurements on jets and plumes with off-source heat addition in a setup similar to that used by BN. Here we consider two flows measured by him: a diabatic jet and a diabatic plume.

### 2.2.1 Diabatic Jet:

**2.2.1-(a) Centreline Velocity ($U_c$)**

The centreline velocity as extracted from figure 3.15a from $V_T$ is plotted in figure 5. It was used for calculating $\alpha_E$ without any further refinement/smoothing.

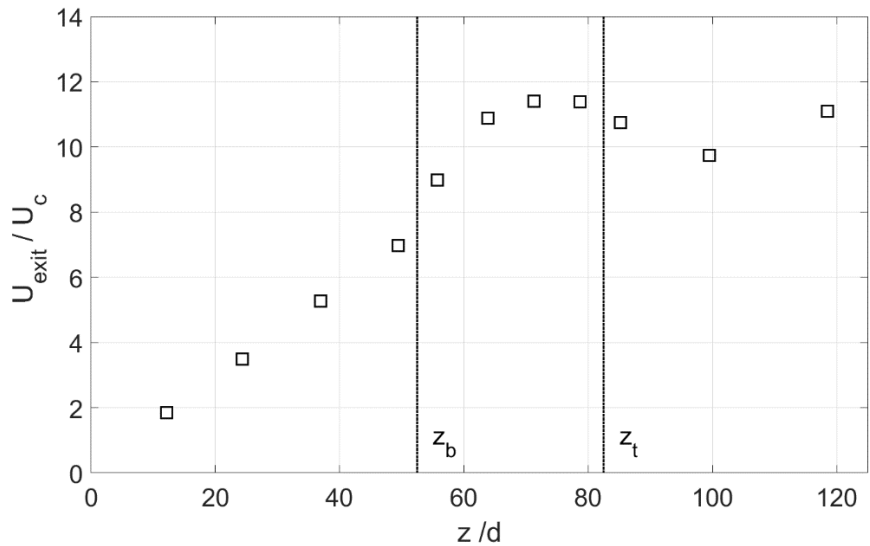

Figure 5 Centreline velocity values extracted from $V_T$ (figure 3.15a) for the diabatic jet. $U_{exit}$ is the orifice exit velocity.

Note that $U_{exit}$ used here is the same as $U_o$ used earlier in the report, denoting orifice exit velocity. We prefer to retain the same notation as used in $V_T$ so that a direct comparison can be made with the plots reported by him.

### 2.2.1-(b) Velocity Width ($b_u$)

$V_T$ has given velocity widths ($b_u$) with $U(b_u) = U_c/2$, presumably because it is more difficult to measure $b_{ue}$ accurately as compared to $b_u$. This data as extracted from figure 3.18 from $V_T$ is plotted in figure 6. As with $U_c$, the velocity width variation was used for calculating $\alpha_E$ without any further refinement/smoothing.

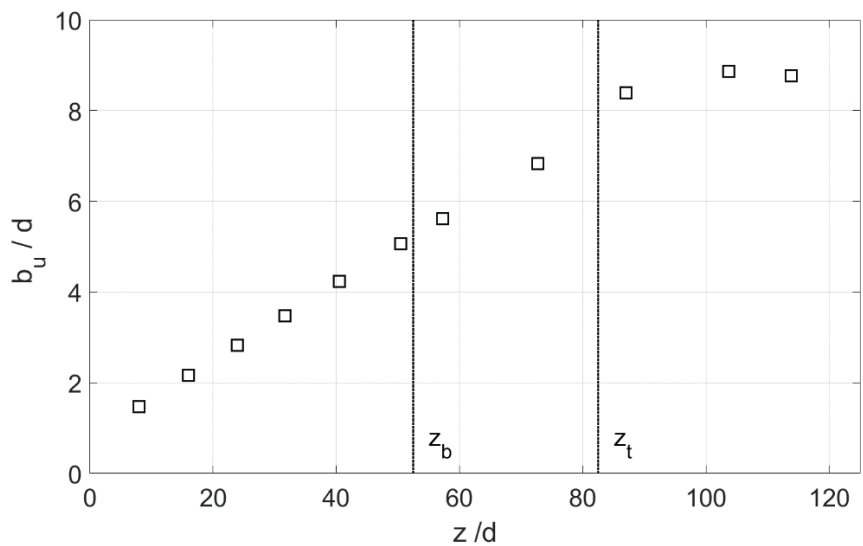

Figure 6. Velocity width values extracted from $V_T$ (figure 3.18) for the diabatic jet.

### 2.2.1-(c) Mass-flow rate ($m$)

$V_T$ has given values of integrated mass-flow rate (per unit density, $m/\rho$) by directly evaluating the integral $\int_0^\infty 2\pi U r dr$ and they are reproduced (from figure 3.19 in $V_T$) in figure 7 below. This is the case for both the jet and the plume.

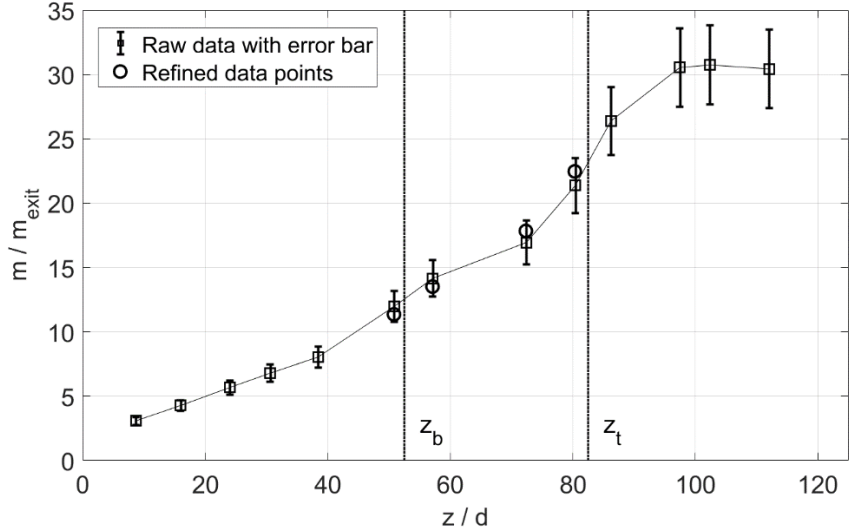

Figure 7 Mass flow rate as a function of $z$ from figure 3.19 in $V_T$ for the diabatic jet. For details see the accompanying text.

It is evident from figure 7 that the measured mass-flow-rate values inside the HIZ i.e. $z_b < z < z_t$, apparently show a non-smooth variation due to the measurement uncertainties. This is particularly seen in the neighbourhood of $z/d \sim 70$. Since differentiation further amplifies small variations, it was thought necessary to refine the data close to and inside the HIZ. This requires estimates of the measurement uncertainty which are not given in $V_T$. However, Venkatakrishnan *et al.* (1999) have given measurement uncertainties in $U_c$ and $b_{ue}$ for similar experiments done using the same setup as used by $V_T$. From these estimates, measurement uncertainty in mass-flow rate is taken to be $\pm 10\%$ of the measured value (see Appendix A for more details). This is shown in figure 7 in terms of error bars on the measured values (square symbols). In order to make sure that the gradients change less abruptly (and therefore are smoother and more realistic) close to and inside the HIZ, refined values of mass-flow rate are selected within the error bar (away from the measured value by approximately 5% on the appropriate side; 5% being half the one-sided error of 10%). These are shown by circles in figure 7.

## 2.2.1-(d) Entrainment Coefficient ($\alpha_E$)

The entrainment coefficients calculated from equation (2.1) using half-velocity width ($b_u$) are plotted in figure 8 for the raw and refined values of mass-flow rate (see figure 7) for the jet. For this (and all the following cases) shape-preserving spline interpolation has been used to get smoother variation of the mass-flow-rate derivative (*dm/dz*). Note that with $b = b_u$, $\alpha_E = 0.065 - 0.068$ for classical round jets (see Appendix B for details). Figure 8 shows that within the measurement uncertainty (indicated by error bars; see Appendix A), $\alpha_E$ reaches a constant value for a small distance upstream of the beginning of HIZ and the range $\alpha_E = 0.065 - 0.068$ falls within this band. The jet entering the HIZ can thus be regarded as nearly self-similar. The linear variation of centreline velocity and width seen from figures 5 and 6 clearly support this conclusion.

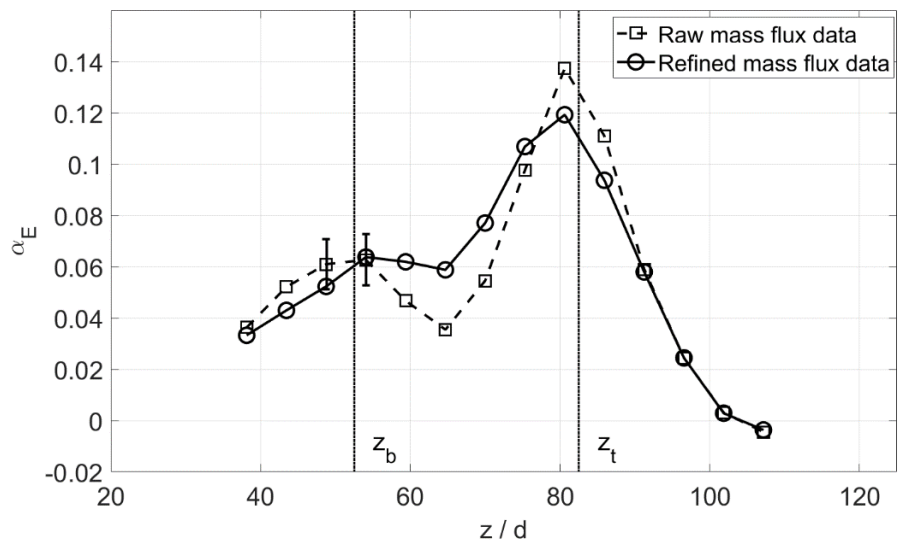

Figure 8. Entrainment coefficient as a function of *z* for the data in figures 5, 6 and 7 (for both raw and refined values) from the measurements of $V_T$ for the jet.

This exercise shows how a small error in measurement of mass-flow rate can introduce a large variation in values of entrainment coefficient. Apart from the measurement uncertainty, discrete approximation of the derivative operator and data interpolation also contribute to the observed variability of data points in figure 8. Note, however, that these do not affect the overall trend in the variation of $\alpha_E$ with *z*. Also note that the variation of $\alpha_E$ with $z$ seen in figure 8 is qualitatively similar to that obtained for the data of BN (see figure 4).

### 2.2.2 Diabatic Plume:

**2.2.2-(a) Centreline Velocity ($U_c$)**

The centreline velocity as extracted from figure 4.16a from $V_T$ is depicted in figure 9. The estimated error bars are shown on the raw data (see Appendix A for further details). The typical uncertainty in $U_c$ is $\pm 5\%$. Two data points have been refined (by choosing values about 2.5% on the positive side of the error bar) so as to make the trend smoother as shown in the figure (although this does not result in substantial improvement).

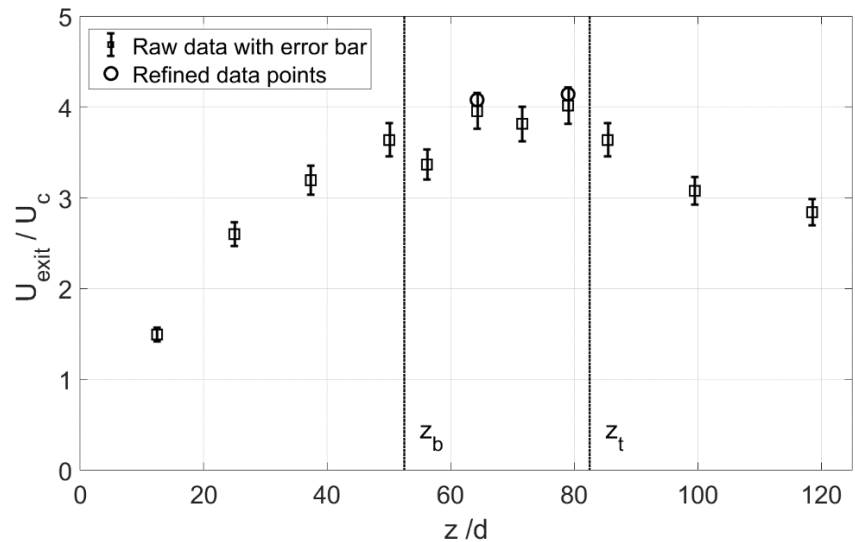

(a) (For caption see the next page)

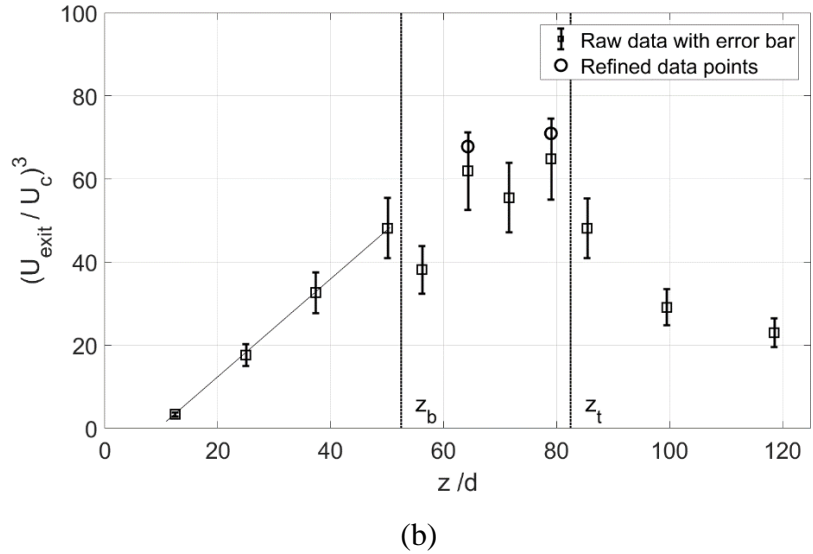

(b)

Figure 9 (a) Centreline velocity values extracted from $V_T$ (figure 4.16a) for the diabatic plume. (b) Centreline velocity plotted as $(U_{exit}/U_c)^3$ v/s z/d; the plume exhibits a linear variation under this scaling upstream of HIZ indicating self-similar behaviour.

### 2.2.2-(b) Velocity Width ($b_u$)

The half-velocity width data ($b_u$) as extracted from figure 4.19 from $V_T$ is depicted in figure 10. It was used for calculating $\alpha_E$ for the diabatic plume without any further refinement.

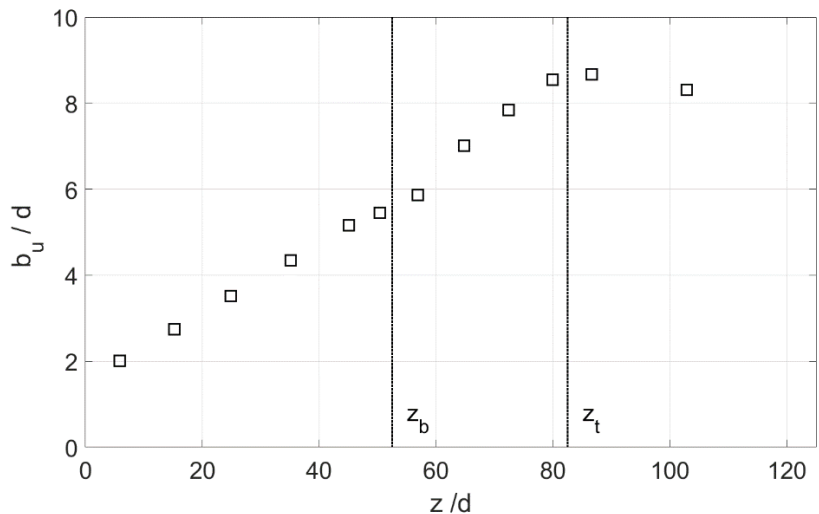

Figure 10 Velocity width values extracted from $V_T$ (figure 4.19) for the diabatic plume.

### 2.2.2-(c) Mass-flow rate (*m*)

The mass-flow rate data ($m/\rho$) for the plume obtained on similar lines as described for the jet is shown in figure 11.

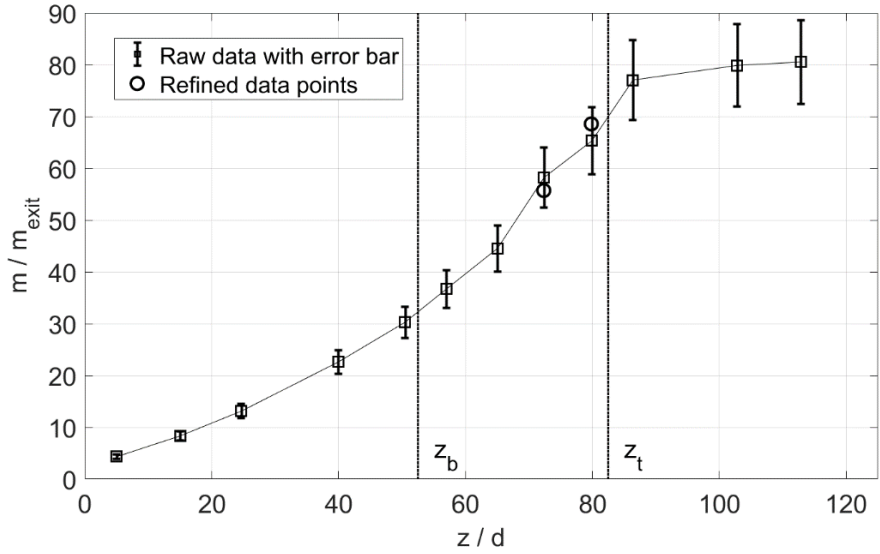

Figure 11 Mass-flow rate as a function of $z$ from figure 4.20 in $V_T$ for the diabatic plume.

### 2.2.2-(d) Entrainment Coefficient ($\alpha_E$)

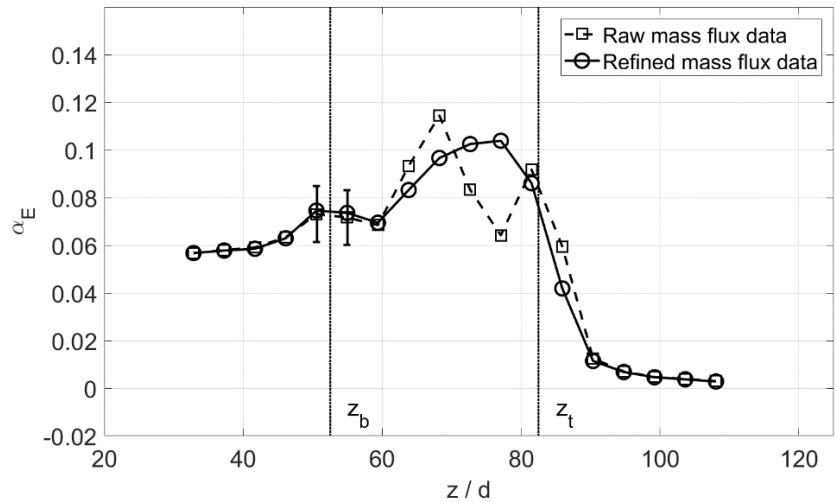

Figure 12 Entrainment coefficient as a function of $z$ for the data in figures 9, 10 and 11 (for both raw and refined values) from the measurements of $V_T$ for the plume.

The entrainment coefficients calculated from equation (2.1) using half-velocity width ($b_u$) are plotted in figure 12 for the raw and refined values of $m$ and $U_c$ (see figures 9 and 11) for the plume.

Again, it is evident that close to the beginning of the HIZ, $\alpha_E$ becomes relatively constant showing the approach to self-similarity. However, the value of $\alpha_E$ (based on $b_u$) is about 20% lower than expected (~0.1; see Appendix B). This could be due to the difficulties encountered in the velocity measurements. It is reasonable to expect that the $\alpha_E$ values will be underestimated at all $z$ locations by more or less the same factor (~20%) since similar procedures were adopted at all the locations. Since we are interested in the variation of $\alpha_E$ relative to its value at $z_b$ (as will be presented at the end of this section), we believe that the qualitative variations in $\alpha_E$ (in the relative sense) will be realistically captured. Again, the sensitivity of calculation of $\alpha_E$ to small variation in mass-flow rate is evident in figure 12. It is interesting to note that in both jet and plume, the data on $\alpha_E$ in figures 4, 8 and 12 show that the nature of the variation of $\alpha_E$ before, within and beyond the HIZ is broadly similar.

## 2.3 Agrawal and Prasad (2004; AP)

AP have performed particle image velocimetry (PIV) measurements on the off-source heated jets. The experimental setup is virtually the same as used by BN. AP did PIV measurements inside the HIZ (where electrode grids are present) using fluorescent particles and a long-wave filter. According to them the scatter seen in the time-averaged data (to be presented below) is due to the presence of the grids and not due to insufficient averaging time.

## 2.3-(a) Centreline Velocity ($U_c$)

The averaged centreline velocity as extracted from figure 8 of AP is depicted in figure 13 below. Since there is a lot of scatter in the data as apparent in figure 13 (which is also true for velocity-width and mass-flow-rate data), it was decided to fair a smooth curve through the data points (solid line in figure 13). The faired curve was drawn by hand; see section 2.3 (c) for the justification.

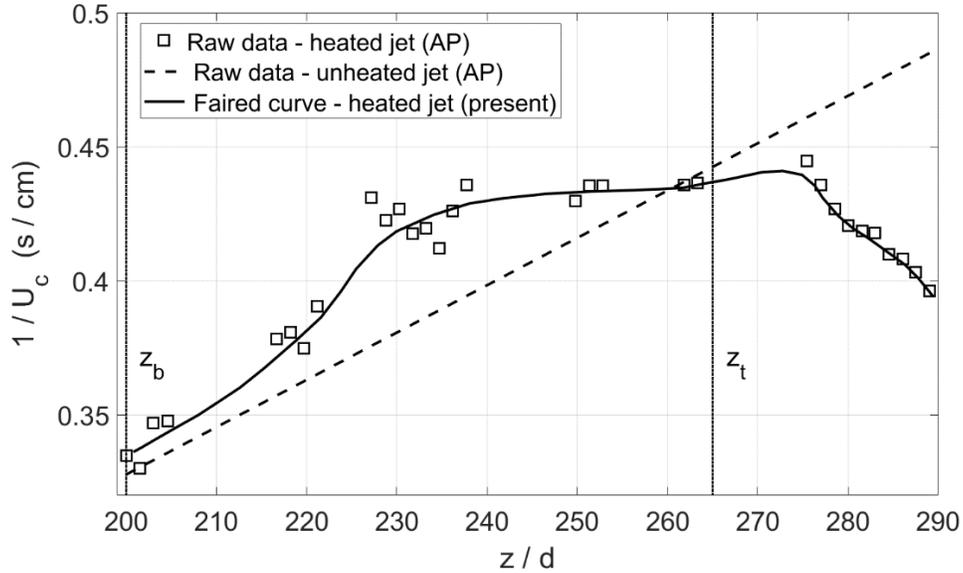

Figure 13 Centreline velocity values extracted from figure 8 in AP along with the faired curve used in the present analysis. $\overline{G} = 4.3$

It is seen that the centreline velocity drops faster than $1/z$ (i.e. that for the unheated jet) in the initial region of the HIZ. The rate of deceleration then decreases, and the centreline velocity starts accelerating a little downstream of $z_t$. AP comment that BN did not report the excess deceleration downstream of $z_b$. Even though BN did not mention it explicitly, it is clear from figure 1 above that the rate of deceleration for their heated jet is slightly higher (than the unheated jet) downstream of $z_b$, and the acceleration begins somewhere *in* the middle of the HIZ. A similar trend is seen in figure 5 ($V_T$-jet) also, except that the acceleration starts close to the end of the HIZ. Thus, apart from the quantitative differences as regards the amount of excess deceleration and the location where the acceleration begins, the qualitative variation seen by AP is similar to that reported in both BN and $V_T$ jets. Note that the observed differences could also be due to differences in the precise distribution of the added heat within the HIZ, which was not measured in any of these studies.

### 2.3-(b) Velocity Width ($b_{ue}$)

The velocity width $b_{ue}$ (where $U(b_{ue}) = U_c / e$) extracted from figure 9 in AP is shown in figure 14 below. Again, the solid line indicates the faired curve. AP have shown a straight line (dash-dot line in figure 14) corresponding to the unheated jet. They compare the variation of $b_{ue}$ for the heated jet with this line and conclude that the velocity width throughout the HIZ exceeds that of a normal unheated jet. Moreover, by linking the variation in $b_{ue}$ with that in $b_{se}$ (scalar width), they find this result to be in contrast with that of BN, wherein $b_{se}$ drops below the corresponding unheated case in the latter portion of the HIZ.

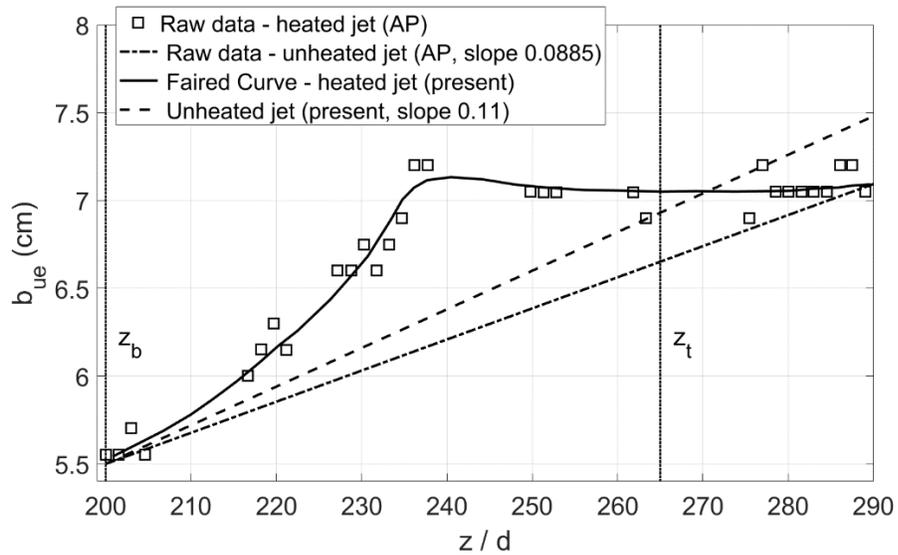

Figure 14 Velocity width values extracted from figure 9 in AP along with the faired curve used in the present analysis. $\overline{G} = 4.3$

We first consider the growth rate for the unheated jet in AP. A closer look at the figure reveals that the slope of the dash-dot line (which is the same as the solid line in figure 9 in AP) is 0.0885. To compare this value for the slope of the unheated jet velocity-width with those of others, we substitute the similarity laws for the axial variation of $U_c$ and $b_{ue}$ for an unheated jet (i.e., $U_c \propto 1/z$ and $b_{ue} \propto z$) into equation (2.1), and obtain $\alpha_E = \frac{1}{2} \frac{db_{ue}}{dz}$. This expression implies that for a typical range of values of $\alpha_E$ reported in the literature for a round jet, i.e., $\alpha_E = 0.054 - 0.0585$ (Turner 1986, Hussain et al. 1994, Bhat and Narasimha 1996), $db_{ue}/dz$ should vary over the range $0.108 – 0.117$. Thus, the slope of 0.0885 for $db_{ue}/dz$ corresponding to the dash-dot line in figure 14 here is clearly too low (by about 20%) as compared to the range $0.108 – 0.117$, found in other well-known data on unheated jets. In this connection Agrawal (2002), in his thesis, has reported $db_{ue}/dz = 0.11$ for the unheated jet for $110 \leq z/d \leq 175$, for the same flow conditions as in AP. (Incidentally this value is consistent with the typical range of $db_{ue}/dz$ obtained from others' data as seen above; see also table 1 below). Making the reasonable assumption that the same slope of 0.11 continues to hold for $z/d > 200$ in an unheated jet, the straight line corresponding to the growth rate of the unheated jet should be close to the dashed line (with slope 0.11) shown in figure 14. Now if we compare the data points for the heated case with the dashed line, we see that the heated values indeed drop below the corresponding unheated values (for $z > z_t$), and the overall variation seen in figure 14 with respect to the dashed line is qualitatively similar to that in figure 2 above (i.e., BN's data and also figure 6 for $b_{se}$ in BN). The only difference is the $z$-location beyond which the heated values of $b_{ue}$ drop below the corresponding unheated values.

It is interesting to seek the source of the slope 0.0885 as used by AP in their figure 9 (and figure 14 here). For a Gaussian velocity distribution, it is shown in appendix B that $b_{ue}/b_u = 1.20$, $b_u$ being the half-velocity width. This gives $db_u/dz = 0.11/1.2 \approx 0.09$. This number is close to 0.0885, which may be compared with the values of the rate of growth of the half-velocity radial width, $db_u/dz$, obtained by other investigators: 0.094 (Hussain et al. 1994), 0.086 (Wygnanski and Fiedler 1969) and 0.09 (Bhat and Narasimha 1996). This suggests the possibility that AP used 'half-velocity' width growth of the unheated jet to compare with '1/e-velocity' width growth of the heated jet. If this were true, it means that they have based their conclusions and the contrasting behaviour of their results with those of BN on this incorrect comparison. Table 1 compares the entrainment coefficient and slopes of the velocity widths (based on half velocity, $b_u$ and 1/e velocity, $b_{ue}$) obtained by various investigators for a classical unheated round jet.

| Investigators | $\alpha_E$ | $db_u/dz$ | $db_{ue}/dz$ |
|---|---|---|---|
| Agrawal and Prasad (2004)[#] | 0.055 | (0.0885) | 0.0885 (0.11) |
| Agrawal (2002) | - | - | 0.11 |
| Wygnanski and Fiedler (1969) | - | 0.086 | - |
| Fisher et al. (1979) (from Turner 1986) | 0.054 | - | 0.108 |
| Hussain et al. (1994) | 0.057 | 0.094 | 0.114 |
| Bhat and Narasimha (1996) | 0.0585 | 0.09 | 0.117 |

Table 1. Summary of the entrainment coefficient and the velocity widths for classical unheated jets obtained by various investigators. Here $\alpha_E$ is based on $b_{ue}$, and $db_{ue}/dz$ is obtained using the relation $\alpha_E = \frac{1}{2}\frac{db_{ue}}{dz}$. Note that in the case of Hussain et al. (1994) the entrainment coefficient ($\alpha_{TH}$) is defined with respect to 'top-hat' variables; they reported $\alpha_{TH} = 0.081$. To obtain $\alpha_E$ as defined in equation (2.1) we have converted the top-hat variables to those relevant for the Gaussian velocity distribution. This gives $\alpha_E = \alpha_{TH}/\sqrt{2}$. [#] $db_{ue}/dz = 0.0885$ as obtained from figure 9 in AP (and figure 14 in this report) should really be $db_u/dz$ whereas $db_{ue}/dz$ is likely to be ~ 0.11 as inferred from Agrawal (2002). The more likely values for $db_u/dz$ and $db_{ue}/dz$ for AP are shown in brackets; see the adjoining text for more details.

**2.3-(c) Mass-flow rate (*m*)**

The mass-flow-rate data ($m/\rho$) extracted from figure 14 in AP is shown here in figure 15 as squares. AP mention that they calculated mass-flow rate using the formula $\pi b^2 U_c$, which is exact for a Gaussian velocity profile. We calculated mass-flow rate using the same formula and using the raw velocity and width data from figures 13 and 14 (for the heated jet of AP), which is shown in figure 15 as triangles. These two data sets do not show a precise match, which is somewhat unexpected. Especially for $230 \leq z/d \leq 280$, the reported mass-flow rate values (squares) are seen to be higher than those calculated from velocity and width (triangles) using the Gaussian assumption. AP discuss briefly about the velocity profile in the HIZ being a flat-topped Gaussian and therefore the actual mass-flow rate being higher than that obtained by $\pi b^2 U_c$. However, it is not clear whether they applied any correction to the calculated values of the mass-flow rate; they do not report any correction having been applied in this regard.

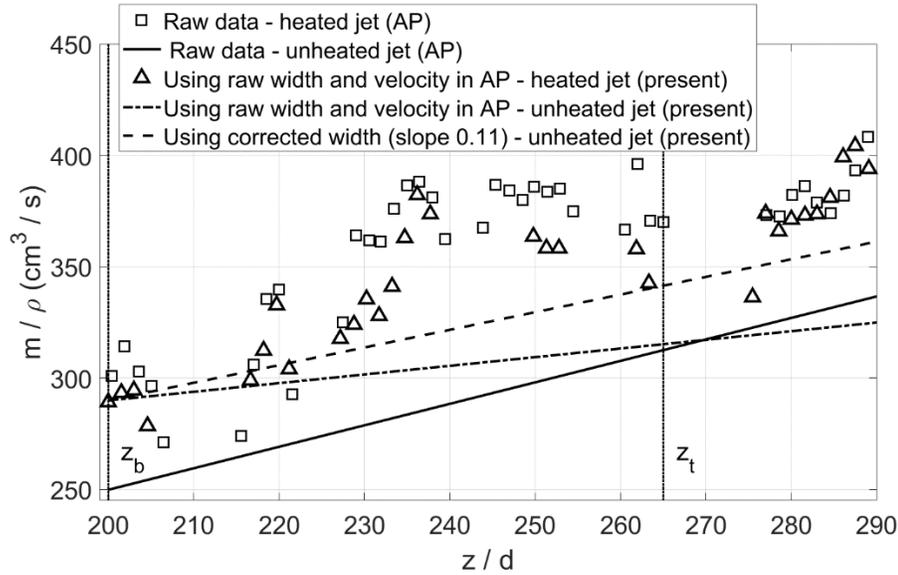

Figure 15 Mass-flow rate as a function of $z$ from figure 14 in AP along with the mass-flow rate calculated for the heated and unheated jets in the present analysis. $\overline{G} = 4.3$

The solid line in figure 15 is the mass-flow-rate variation for the normal (unheated) jet as given in figure 14 in AP. The dash-dot line is obtained by calculating mass-flow rate for the unheated jet using the unheated data reported by AP for velocity and width from figures 13 (dashed line) and 14 (dash-dot line) above. Again, these two lines do not coincide for some reason. In fact, the values for the unheated jet at $z/d = 200$, taken from figures 13 and 14, are $1/U_c = 0.3277 \ s/cm$, i.e., $U_c = 3.05 \ cm/s$ and $b_{ue} = 5.5 \ cm$. This gives a mass-flow rate of $289.8 \ cm^3/s$ but the unheated mass-flow rate reported by AP is 250 $cm^3/s$ (as seen in figure 15). The reasons for this discrepancy are not clear. Finally, the dashed line in figure 15 shows the unheated mass-flow rate using the corrected unheated width, with slope 0.11, from figure 14 (dashed line). In the following, we have chosen the dashed line in figure 15 to represent the variation of mass-flow rate for the unheated jet, as this seems to be the most consistent variation in view of the above discussion.

For drawing a faired curve through the mass-flow-rate data we have chosen the values directly reported by AP, i.e., the squares in figure 15. Since the scatter in the data is large, more than one choice of smoothed curve through the data is possible. Figure 16 shows two such faired curves which more or less represent the extreme choices, and they were selected so as to represent the possible variability in the calculated values of the entrainment coefficient due to the data scatter. In choosing these curves we were guided by the mass-flow-rate variation for the unheated jet; the heated mass-flow rate departs from the unheated value close to $z_b$ and shows a sharp rise followed by a weaker variation. This is consistent with the description in AP and with the general trend seen in figures 3, 7 and 11 above.

Note that the faired curves in figures 13, 14 and 16 have been drawn by hand using visual judgement for the best fit. We tried using least square polynomial fits, but owing to the large scatter in the data they produced spurious oscillations especially for the mass-flow-rate data. Since calculation of $\alpha_E$ involves taking derivatives of the mass-flow rate, the fitted data produced unrealistic variations. As a result, curves faired by hand were thought to be more reliable in revealing the trend and variability in $\alpha_E$ and therefore were selected for the present analysis.

The above considerations show that the overall tends in $U_c$, $b_{ue}$ and $m$ in AP are qualitatively similar to those in BN and V$_T$ (except for the dip in the mass-flow rate in AP as in figure 16) and not in contradiction as AP have contended.

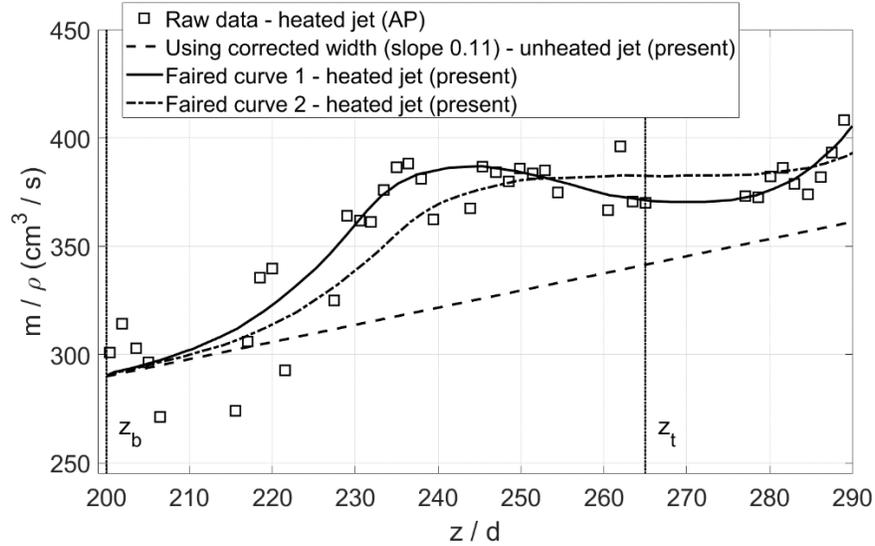

Figure 16 Mass-flow rate as a function of $z$ (raw data from figure 14 in AP) with two choices for the faired curve used for calculating entrainment coefficient in the present analysis. They represent, more or less, the extreme choices for the mass-flow rate variation. See figure 15 for more details.

### 2.3-(d) Entrainment Coefficient ($\alpha_E$)

The variation of $\alpha_E$ with $z$ is plotted in figure 17 below for the two choices of the faired curves for the mass-flow rate shown in figure 16. For $U_c$ and $b_{ue}$, the faired curves shown in figures 13 and 14 respectively were used. The extrapolated values of $\alpha_E$ at $z_b$ for the faired curves 1 and 2 come out to be 0.0554 and 0.048 respectively. These values are in the same ball park as the standard values i.e. 0.054 - 0.0585 (see table 1) indicating the overall soundness of the procedure followed here.

It is seen that for the choice of curve 1 in figure 16, $\alpha_E$ values become negative towards the end of HIZ and show a sharper rise for $z > z_t$ as compared to curve 2.

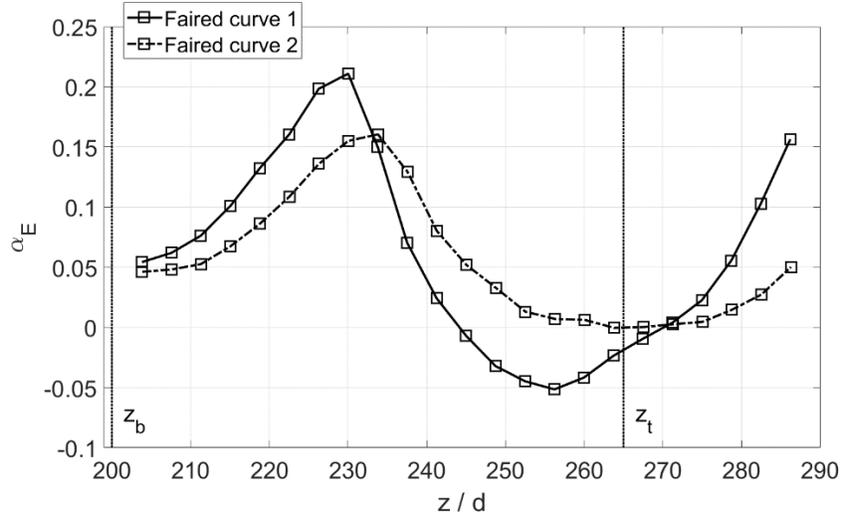

Figure 17 Entrainment coefficient as a function of $z$ for the data in figures 13, 14 and 16 from the measurements of AP.

## 2.4 L. Venkatakrishnan, R. Elavarasan, G. S. Bhat, A. Krothapalli and L. Lourenco (2003; VK)

VK performed PIV measurements on a jet with off-source heating in a setup similar to that used by BN. They have reported the values of the entrainment coefficient by directly measuring the radially-inward velocity at the jet edge ($b_{ue}$) in an axial section of the flow. The $\alpha_E$ values extracted from figure 7 from VK are reproduced here in figure 18. They have not reported any measurements inside the HIZ, presumably due to difficulties associated with the presence of heater grids.

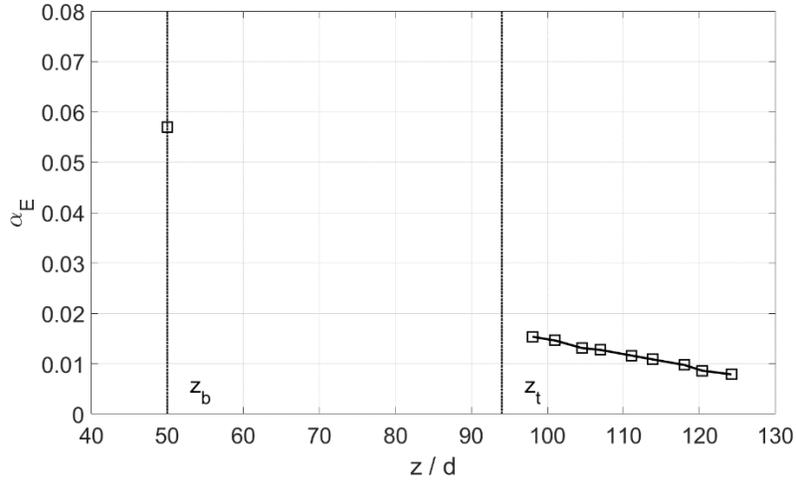

Figure 18 Variation of $\alpha_E$ with $z$ from figure 7 in VK.

VK obtained a value of $\alpha_E = 0.057$ for the unheated jet which matches well with the values reported in the literature (i.e. 0.054 – 0.0585; see table 1).

## 2.5 Summary plot

Figure 19 below is a summary plot showing the entrainment coefficient values from the five data sets mentioned above. Note that the data in Venkatakrishnan *et al.* (1999) has not been included here since the experimental conditions therein were broadly similar to those in $V_T$. Also, $V_T$ carried out experiments both on a diabatic jet and plume, and therefore was chosen for the present analysis.

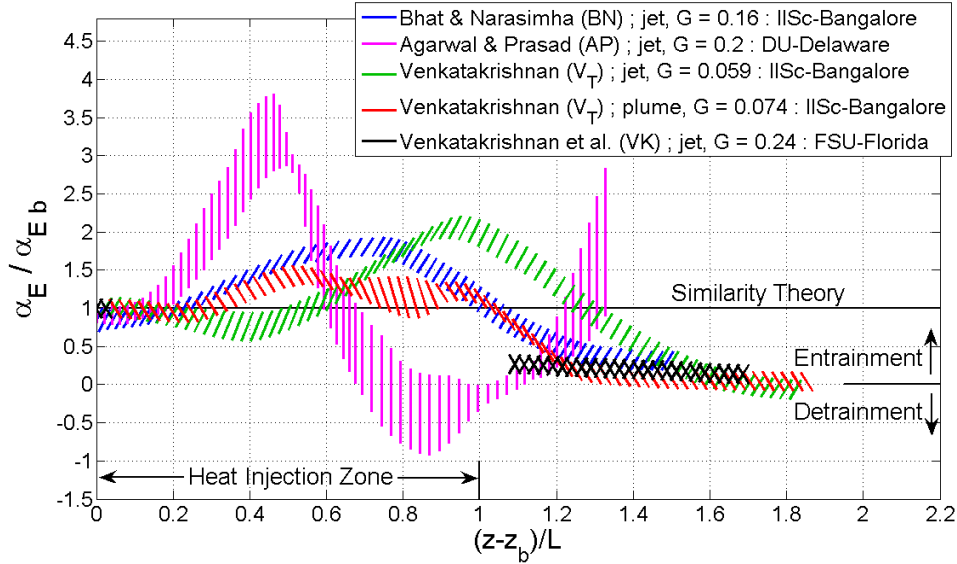

Figure 19 Summary plot of variation of $\alpha_E$ with $z$ for the five data sets discussed above (reproduced from Narasimha *et al.* 2011). For the symbols see the accompanying text.

Here $\alpha_{Eb}$ is the entrainment coefficient at $z = z_b$ and $L$ is the height of the HIZ. The heat-release number $G$ included in figure 19 is defined as follows.

$$G = \frac{\beta_T g}{\rho C_p} \frac{Q}{b_b U_b^3},  \qquad (2.3)$$

where $U_{cb}$ is the centreline velocity at the base of the HIZ. For BN, AP and VK, $b = b_{ue}$ and for $V_T$ (jet and plume), $b = b_u$. However, since we are interested in the relative increase in $\alpha_E$ over its value at $z_b$ (i.e. the ratio $\alpha_E/\alpha_{Eb}$), the choice of the definition of $b$ becomes irrelevant.

Each curve in figure 19 is plotted in the form of a band which indicates a typical variability of the entrainment coefficient due to different choices of curves used for fairing the data as discussed earlier (figures 4, 8, 12, 17). Wherever variability was seen to be small, a band of certain minimum width (only representative and therefore more or less arbitrary) has been added to show the general uncertainty of these values due to that in the measurements (see, e.g., appendix A).

It can be seen that the curves in figure 19 are not universal; their precise variation will depend upon local $G$ and its vertical profile. However, there is a general agreement on the qualitative nature of the effect of off-source heating: an initial rise, followed by a drop to near-zero – possibly even to negative values.

Note that all the values of *G* in figure 19 are with the '1/e-velocity' width $b_{ue}$. For this purpose, for the data-sets of V$_T$, $b_u$ was multiplied by 1.2 to get $b_{ue}$ (which is valid since the measured velocity profiles were approximately Gaussian). Another point we would like to note here is that for the experiments of AP, the two definitions of the heat-release number (equations 2.2 and 2.3) are related as $\overline{G} = 21.6\,G$. However, AP have reported the ratio between $\overline{G}$ and G to be 12.5. (Note that they use different symbols for $\overline{G}$ and G from the ones used here). There seems to have been a numerical mistake in calculating this ratio in AP.

### 3. Laboratory experiments vis-à-vis steady-state deep convection

Romps and Kuang (2010) performed a numerical simulation of steady-state deep convection over tropical oceans using a fully compressible Cloud-Resolving Model (CRM). They defined a quantity called 'purity' (*p*) which is related to the dilution of the cloudy air due to the entrainment of the ambient air. N+ introduced the concept of a 'diabatic purity' ($p_d$) for laboratory diabatic jets and plumes, which can be considered as an analogue of purity for real clouds. They showed a favourable comparison between the diabatic purity for the laboratory flows with the purity variation given in Romps and Kuang (2010). N+ have given a brief derivation of $p_d$ (in their '*Materials and Methods*' section) which finally reads as $p_d(z) = m(z_b)/m(z)$. Here (in section 3.1) we present a more detailed account of the derivation to arrive at the above expression. In section 3.2 we point out the differences in conditions between the laboratory experiments and CRM computations and how they can be reconciled.

## 3.1 Derivation of laboratory diabatic purity

The mass-flux-weighted average concentration of dye particles in the laboratory flows is defined as (N+; *Materials and Methods*)

$$\bar{c}_a(z) = \int_A \left(\overline{c\, \dot{m}_c}\right) dA \bigg/ \int_A \overline{\dot{m}_c}\, dA, \quad (3.1)$$

where $c$ is local concentration, $A$ is cloud cross-sectional area and $\dot{m}_c$ is the laboratory equivalent of the 'cloudy' mass flux given by $\dot{m}_c = k(\Delta T)\dot{m}$; $k$ is the 'activity operator' (see Romps 2010) that differentiates cloudy parcels from the ambient ones ($k=1$ for cloudy parcels and $0$ otherwise), $\Delta T$ is the excess temperature in the diabatically-heated flow over its unheated counterpart. Note that in what follows we use an overbar to denote the time averages.

Substituting for $\dot{m}_c$ in equation (3.1) we get

$$\bar{c}_a(z) = \int_A \left(\overline{c\, k\dot{m}}\right) dA \bigg/ \int_A \left(\overline{k\dot{m}}\right) dA. \quad (3.2)$$

The total measured mass-flow rate and the total 'cloudy' mass-flow rate at a given height $z$ can be related as follows.

$$\int_A \left(\overline{k\dot{m}}\right) dA = \bar{k}^*(z) \int_A \overline{\dot{m}}\, dA, \quad (3.3)$$

where $\bar{k}^*(z)$ may be called the flux-weighted average activity operator: $\bar{k}^*(z) = 0$ for $z \leq z_b$ and takes value between 0 and 1 for $z > z_b$, based on the variation of scalar-concentration width with height.

Since we are dealing with integral quantities for steady flows, it is convenient to use the top-hat formalism. For thin round steady-state jets and plumes that are axisymmetric in the mean, the top-hat profiles can be defined as follows (see Turner 1973),

$$\int_A \overline{\dot{m}}\, dA = \rho \pi R^2 W; \quad \int_A \overline{\dot{m}\, \bar{u}_z}\, dA = \rho \pi R^2 W^2; \quad \int_A \overline{\dot{m}\, \bar{c}}\, dA = \rho \pi R^2 BW, \quad (3.4)$$

where $\bar{u}_z$ is the mean axial velocity, and $R$, $W$ and $B$ are respectively the top-hat radius, vertical velocity and concentration as depicted graphically below.

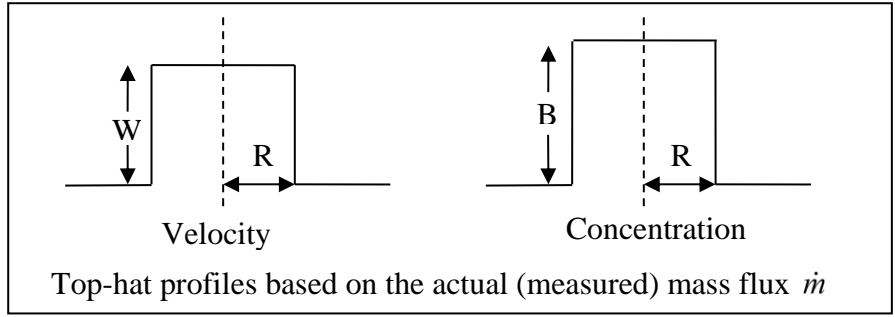

Top-hat profiles based on the actual (measured) mass flux $\dot{m}$

Next we assume

$$\int_A \overline{(k\dot{m})}\, dA \cong \int_A \bar{k}\,\overline{\dot{m}}\, dA, \qquad (3.5)$$

which follows from the thin-shear-layer theory, and is consistent with the neglect of the turbulent transport of concentration already made in N+ (i.e., in arriving at equation 3.1). Using equation (3.4) in equation (3.3),

$$\int_A \overline{k\dot{m}}\, dA = \int_A \bar{k}\,\overline{\dot{m}}\, dA = \bar{k}^*(z)\int_A \overline{\dot{m}}\, dA = \rho\pi\bar{k}^* R^2 W \qquad (3.6)$$

As noted in N+, values of $\dot{m}$ and $\Delta T$ in laboratory flows are, on an average, observed to be higher close to the centreline as compared to those near the edge of the jet/plume. This is so for sufficiently tall cumulus clouds as well (see e.g., Reuter and Yau 1987, Blyth *et al.* 1988). Since the activity operator *k* depends on a threshold on $\Delta T$, it is highly likely that the fluid parcels close to the plume edge have much lower values of *k* as compared to those near the plume core. For the statistically stationary flows we are dealing with here, it is reasonable to expect that there is a certain 'similarity' in the radial distribution of parcels, i.e., the probability of finding fluid parcels with large values of $\dot{m}$ and $\Delta T$, at any time *t*, is always higher close to the plume centreline than near its edge. In other words the 'uncloudy' (or ambient) parcels are more likely to be found near the edge of the plume than its core. Therefore, the action of $\bar{k}$ (taken to be axisymmetric) is to reduce the effective width of the diabatically heated flow that can be considered as cloudy.

With this premise we write (from equation 3.6),

$$\int_A \bar{k}\,\overline{\dot{m}}\, dA = \rho\pi R'^2 W,$$

where $R' = \sqrt{\bar{k}^*}\, R$ is the radius of the top-hat profiles based on the 'cloudy' mass flux, with $R' \leq R$. Thus we can define cloudy top-hat relations as follows.

$$\int_A \bar{k}\,\bar{m}\,dA = \rho \pi R'^2 W \;;\quad \int_A \bar{k}\,\bar{m}\,\bar{u}_z\,dA = \rho \pi R'^2 W^2 \;;\quad \int_A \bar{k}\,\bar{m}\,\bar{c}\,dA = \rho \pi R'^2 BW \;, \qquad (3.7)$$

Here we have taken $W$ and $B$ for the cloudy top-hat profiles to be the same as those for the actual top-hat profiles (equation 3.4), and reduced the radius. This is a reasonable assumption since $\bar{k}(r,z)$ is a binary function (1 or 0) so it does not alter the mass flux of parcels classified as cloudy but only excludes the mass flux of parcels not termed cloudy. Therefore its primary effect is to alter the effective width of the flow and leave the height of the top-hat profiles relatively unchanged.

From equations (3.7) and (3.4),

$$\int_A \bar{k}\,\bar{m}\,\bar{c}\,dA = \bar{k}^* \rho \pi R^2 BW = \bar{k}^* \int_A \bar{m}\,\bar{c}\,dA \qquad (3.8)$$

From equations (3.2), (3.3), (3.5) and (3.8),

$$\bar{c}_a(z) = \frac{\bar{k}^* \int_A \bar{c}\,\bar{m}\,dA}{\bar{k}^* \int_A \bar{m}\,dA} = \frac{\int_A \bar{c}\,\bar{m}\,dA}{\int_A \bar{m}\,dA}$$

Thus, for the conditions mentioned above, precise determination of $\bar{k}^*(z)$ is not necessary for calculating the flux-weighted average concentration for the laboratory flows.

Following the derivation given in N+ after this step, we get the final expression for the diabatic purity as

$$\bar{p}_d(z) = \frac{\bar{c}_a(z)}{\bar{c}_a(z_b)} = \frac{\bar{m}(z_b)}{\bar{m}(z)} \;.$$

### 3.2 Differences in conditions between the CRM computations and laboratory flows.

Note that there are certain differences in conditions between the CRM computations of Romps and Kuang (2010; RK) and the present laboratory simulations. These are as follows.

1. RK use Radiation-Convection Equilibrium condition which typically involves a system of many clouds, whereas the laboratory experiments deal with a single cloud.
2. RK add sources and sinks of purity tracers to enable accurate calculation of purity in the simulated cloud system. On the other hand in the laboratory experiments there are no sources or sinks of dye concentration.

However, these differences are expected to have only a second-order effect on the purity comparison presented in N+. The reasons are given below.

1. **Multiple Clouds:** It is true that the calculation of average purity in RK involves cloudy parcels from all the clouds present in the domain. However, we can still consider the variation in Fig. 5 in N+ as representative of that in an 'average' individual cloud in the CRM computations for the following reasons.

    (a) RK run their numerical simulation for sufficiently long time so that steady-state deep convection is established everywhere in the domain. Thus all the clouds in the domain can be expected to be qualitatively similar (i.e., statistically stationary and deep).

    (b) They take sufficient care to make sure that the entraining air in each convective cloud has zero purity and all the sub-cloud (i.e., below cloud base) air has purity equal to unity. This implies that all the clouds essentially experience a similar environment and undergo dilution in a similar manner.

2. **Sources and sinks of purity:** RK have not given any quantitative information on the strengths of sources and sinks of the purity tracers. They use sinks in the environment (i.e., ambient) where main cloudy updrafts are not present, and sources below the cloud base. In both regions convection is much weaker than in the cloudy updrafts. Since the weighting function used in calculating average purity is the updraft mass flux, which takes small values in both these regions (in fact it is zero below the cloud base), the contribution of purity sources and sinks to the average purity can be considered to be very small. Thus the total transport of purity tracers by the cloudy mass flux in the computations of RK is expected to behave very nearly as a conserved quantity, making it similar to the laboratory experiments.

**4. Homogeneous vs. inhomogeneous mixing in cumulus clouds**

A nondimensional group that distinguishes between the so-called homogeneous and inhomogeneous mixing is the Damkohler number (Da) given by

$$\mathrm{Da} = \frac{\tau_{mix}}{\tau_{react}}.$$

If Da < 1(> 1), the mixing is termed as homogeneous (inhomogeneous). Here $\tau_{mix}$ is the turbulent mixing time scale and $\tau_{react}$ is the 'reaction' time scale associated with the phase change, i.e., droplet evaporation in the present context (Lehmann *et al.* 2009). The mixing time scale is given by $\tau_{mix} = \left(l_E^2/\varepsilon\right)^{1/3}$, where $l_E$ is the length scale of the entrained parcels, and $\varepsilon$ is the turbulent kinetic energy dissipation rate.

N+ have reported the estimates of the turbulent mixing time scale in the field measurements on cumulus clouds by Gerber *et al.* (2008; to be called Gerber+ in what follows) and the laboratory experiments on diabatic jets by Bhat and Narasimha (1996). They observed that the mixing tends to become more homogeneous as we move up from the cloud base, in both natural and laboratory clouds. N+ have given a brief account of how the estimates of $\tau_{mix}$ were made; see section 6 in the Supporting Information in their paper. In this section we provide additional information in support of the arguments made therein.

The mixing time scale in the clouds measured by Gerber+ varies from $\tau_{mix} \approx 12$ s at level 1 (252m above cloud base) to $\tau_{mix} \approx 7$ s at level 4 (918m above cloud base); see N+. This means that the turbulence in these clouds tends to enhance mixing with height above the cloud base. However, to conclude whether the mixing is homogenous or not requires values of $\tau_{react}$ also (to be able to calculate the Damkohler number). Unfortunately, these are not available in the measurements of Gerber+. In this connection, Lehmann *et al.* (2009) have given some estimates of $\tau_{react}$ for idealized uniform-size droplet populations. For the typical values of the droplet number density (~100 cm$^{-3}$) and diameters (~20 μm) found in the measurements of Gerber+, $\tau_{react}$ is seen to increase with increase in the ambient saturation ratio, i.e., increase in height (see Fig. 2 in Lehmann *et al.* 2009). Thus, the combined effect of decrease in $\tau_{mix}$ and increase in $\tau_{react}$ is to decrease the Damkohler number with height, thereby making the mixing more homogeneous. This is consistent with the conclusion in Gerber+**,** which they arrive at based on similar considerations (page 98 in their paper) and supports the scenario presented in N+. Note that for certain combinations of the droplet number density and diameter, $\tau_{react}$ is seen to decrease mildly with height (figure 2 in Lehmann *et al.* 2009). In such cases the rate of decrease of $\tau_{mix}$ with height in relation to that of $\tau_{react}$ will decide the nature of mixing. In any case, what is clear is that the turbulence structure of cumulus clouds will *tend* to make mixing more homogeneous with height (provided we are not too close to the cloud top).

Curiously, the mixing diagrams plotted in Gerber+ present a somewhat contradictory picture, i.e. mixing becoming more inhomogeneous with height. However, there are certain ambiguities in interpreting the mixing diagram as recognized by Gerber+ themselves. It is not possible to distinguish between inhomogeneous and homogeneous mixing when the ambient relative humidity is close to unity, and it does approach unity as the height is increased. Also secondary activation of cloud droplets due to entrained condensation nuclei could give a false indication of mixing becoming more homogeneous (see Fig. 8 in Gerber+ and the adjoining discussion). Furthermore, it is clear from the measurements of Lehmann *et al.* (2009) that in the regions of increased dissipation inside the clouds the mixing is more homogeneous. Therefore, notwithstanding the (somewhat ambiguous) interpretations of the mixing diagrams, it is reasonable to conclude that mixing tends to become more homogeneous with height in the shallow cumulus clouds measured by Gerber+.

Next we consider the laboratory experiments on steady diabatic jets by Bhat and Narasimha (1996) and provide additional supporting information with regard to the choice of length scales made in calculating the mixing time scales. Table 2 shows values of various parameters at two heights - one at the base of the HIZ and the second just above the end of the HIZ. The mixing time scale is calculated as $\tau_{mix} = \left(l_E^2 l/u^3\right)^{1/3}$ (N+), where $u$ and $l$ are respectively the typical large-eddy velocity and time scales.

| $(z - z_b)/L$ | $b$ (m) | $u$ (m/s) | $l$ (m) | $l_E$ (m) | $\tau_{mix}$ (s) |
|---|---|---|---|---|---|
| 0 | 0.025 | 0.0108 | 0.025 | 0.0125 | 1.5 |
| 1.15 | 0.036 | 0.0086 | 0.018 | 0.0036 | 0.7 |

Table 2. Calculation of $\tau_{mix}$ in the laboratory measurements of Bhat and Narasimha (1996). Reproduced from N+.

N+ have reported following estimates for the length scales used in calculating $\tau_{mix}$. At $(z-z_b)/L = 0$, $l = b$ and $l_E = b/2$; at $(z-z_b)/L = 1.15$, $l = b/2$ and $l_E = b/10$. These are mainly obtained by a close examination of the instantaneous flow-visualization pictures of the planar sections of these flows. The large-eddy length scale (*l*) is estimated by visually identifying a typical large eddy based on the regions of enhanced dye concentration, whereas the entraining parcel length scale (*l$_E$*) is estimated by identifying regions greatly depleted in dye concentration. The choice of *l* is consistent with the wavelet analysis of the diametral sections of the flow (Narasimha *et al.* 2002), wherein at an appropriate wavelet scale the shape and size of the coherent structures are revealed. Figure 3f in Narasimha *et al.* (2002) shows that the length scale of the coherent structure is of the order of the local flow width for the unheated plume, and it is roughly half of the width for the heated plume. Moreover, since a relatively well-mixed protected core is seen to be present in a diabatic plume, the action of large eddies is likely to be limited primarily to a ring surrounding the core. This implies that the large-eddy length scale for a diabatic jet/plume would be a fraction of that for an unheated jet/plume. A further support for the choices of *l$_E$* in table 2 comes from the direct numerical simulations of a temporally growing jet with off-source heating performed by Basu and Narasimha (1999). Figures 10a,b in their paper show the drastic reduction in the length scales of the regions occupied by the entrained fluid (represented by very low values of vorticity) in the heated jet (figure 10b) as compared to the unheated jet (figure 10a).

The reduced values of *l* and *l$_E$* obtained above the HIZ (see table 2) are due to the disruption of large-scale coherent structures as clearly seen in the flow visualization pictures (see N+). With these estimates, the calculated value of $\tau_{mix}$ in the laboratory experiments decreases from approximately 1.5s at the base of the HIZ to 0.7s above the HIZ. Note that even though there is some subjectivity in choosing *l* and *l$_E$*, especially at $(z-z_b)/L = 1.15$, the qualitative behaviour of the decrease in mixing time with height remains the same. For example, as an extreme choice, if we take $l = 2b/3$ and $l_E = b/5$ at $(z-z_b)/L = 1.15$, we get $\tau_{mix} = 1.25\,\text{s}$, which still lower than $\tau_{mix} = 1.46\,\text{s}$ at $(z-z_b)/L = 0$ (rounded to 1.5 in table 2). Thus the overall qualitative behaviour is independent of the precise choice of the estimates of length scales.

Another observation worth noting is the effect of off-source diabatic heating on the r.m.s. turbulence level ($\hat{u}$). In the absence of heating, the maximum value of $\hat{u}$ at $(z - z_b)/L = 1.15$ measured by Bhat and Narasimha (1996) is equal to $0.25 U_c$ or $0.062$ m/s. Table 1 shows that with off-source heating there is about 25% increase in the turbulence level at the same height; also the peak in the radial distribution of $\hat{u}$ (which is off-centre) gets more pronounced as compared to the centreline value (Fig. 9d in Bhat and Narasimha). The enhancement of small-scale vorticity due to the baroclinic torque (Basu and Narasimha 1999) generated by off-source heating is thought be responsible for this behaviour.

## 5. Summary

In this report we have provided additional supporting information about three entrainment-related issues reported in Narasimha *et al.* (2011; N+) that are relevant for the dynamics of cumulus clouds. They are as follows.

(a) A critical re-analysis of the laboratory data on steady diabatic jets and plumes has been carried out, which has revealed the variation of entrainment coefficient with height. The analysis requires some data smoothing and refinement. Sufficient care has been taken in making such refinements by making consistency checks wherever possible and presenting the reasoning behind each step. In particular, it is found that there is an inconsistency in the use of the measure for the width of the jet in the work of Agrawal and Prasad (2004). One of the main outcomes of this exercise is that, if this inconsistency is removed, the data-set of Agrawal and Prasad is found to be broadly consistent with the other measurements included in the analysis, contrary to what Agrawal and Prasad have concluded. The present exercise clearly brings out the anomalous-entrainment behaviour typical of cumulus clouds (see N+ for more details).

(b) A detailed derivation is presented (in addition to what is given in N+) to arrive at the expression of diabatic purity used in N+. The assumptions made therein have been clearly spelt out. Furthermore, we have listed the differences in conditions in the CRM computations and laboratory experiments, and provided arguments to show that they are likely to have only a second-order effect on the favourable comparison of diabatic purity with that obtained from CRM computations (N+).

(c) Additional information is given in support of the choice of length scales for estimating turbulent mixing time scales in both field and laboratory measurements of cumulus clouds. Additional arguments are provided to support the contention made in N+ that the mixing tends to become more homogeneous as we move up from the cloud base, both in natural clouds and in laboratory simulations.

**Appendix A: Estimates of measurement uncertainty in $m$ and $\alpha_E$ for the jet and plume studied in Venkatkrishnan-1997 ($V_T$)**

As mentioned in the main text, the uncertainty estimates for the measurements reported in $V_T$ are obtained from Venkatakrishnan *et al.* (1999; VBN), wherein measurements from the same setup are reported for a plume under different conditions. VBN have given uncertainty levels for $U_c$ and $b_{ue}$ in figures 4 and 7 respectively in their paper. Typical values of these levels as extracted from VBN are as follows:

$b_{ue}/d = 10.68$, $\delta(b_{ue}/d) = \pm 0.68$ and

$(U_o/U_c)^3 = 82.14$, $\delta[(U_o/U_c)^3] = \pm 12.5$,

where $\delta$ is the uncertainty. As there is little uncertainty in $d$ and $U_o$, we can write the relative uncertainties as follows.

$$\frac{\delta(b_{ue}/d)}{(b_{ue}/d)} = \frac{\delta b_{ue}}{b_{ue}} \text{ and } \frac{\delta[(U_o/U_c)^3]}{(U_o/U_c)^3} = -3\frac{\delta U_c}{U_c}$$

Now the mass-flow rate $m \propto b_{ue}^2 U_c$. Following the standard practice for calculating the propagation of uncertainties, we write

$$\frac{\delta m}{m} = \sqrt{\left(2\frac{\delta b_{ue}}{b_{ue}}\right)^2 + \left(\frac{\delta U_c}{U_c}\right)^2} = \sqrt{\left(2\frac{0.68}{10.68}\right)^2 + \left(\frac{12.5}{3 \times 82.14}\right)^2} \approx 0.137$$

Note that this value of uncertainty is based on the uncertainty in $b_{ue}$, which is expected to be higher than that in $b_u$ because of difficulty in measuring small velocities close to the edge sufficiently accurately. Furthermore, in $V_T$, the mass-flow rate is obtained by integrating the velocity profile in the radial direction. As a result, this uncertainty estimate (0.137) is likely to be an overestimate. To account for this and also considering the fact that the experimental conditions in $V_T$ and VBN are different, we take the relative uncertainty in the mass-flow rate to be $\pm 10\%$. The error bars shown in figures 7 and 11 are drawn with this level of uncertainty. The uncertainty in $\alpha_E$ is calculated the following way.

$$\frac{\delta \alpha_E}{\alpha_E} = \sqrt{\left[\frac{\delta(\Delta m)}{\Delta m}\right]^2 + \left[\frac{\delta b_{ue}}{b_{ue}}\right]^2 + \left[\frac{\delta U_c}{U_c}\right]^2} = \sqrt{\left(\sqrt{2}\times 0.1\right)^2 + \left(\frac{0.68}{10.68}\right)^2 + \left(\frac{12.5}{3\times 82.14}\right)^2} \approx 0.16$$

Here, $\delta(\Delta m)/\Delta m$ is taken to be $\sqrt{2}\,\delta(m)/m$ again in the root mean square sense. Error bars shown in figures 8 and 12 represent an uncertainty level of $\pm 16\%$.

**Appendix B: Entrainment coefficient for classical self-similar jets and plumes based on $b_u$ and $b_{ue}$**

Entrainment coefficients based on $b_u$ and $b_{ue}$ are defined (see equation 2.1) as

$$\alpha_{Eb_u} = \frac{dm/dz}{2\pi\rho b_u U_c} \tag{B1}$$

$$\alpha_{Eb_{ue}} = \frac{dm/dz}{2\pi\rho b_{ue} U_c} \tag{B2}$$

Dividing equation (B1) by (B2),

$$\alpha_{Eb_u} = \frac{b_{ue}}{b_u} \alpha_{Eb_{ue}} \tag{B3}$$

To obtain the relation between $b_u$ and $b_{ue}$, we make use of the observation that in both round jets and plumes the axial velocity distribution in the radial direction ($r$) is very nearly Gaussian (see Turner 1986), i.e.,

$$\frac{U}{U_c} = \exp\left(-\left(\frac{r}{b_{ue}}\right)^2\right) \tag{B4}$$

Using the definition of half-velocity width, $r = b_u$ when $U = U_c/2$. Putting this in equation (B4),

$$\frac{1}{2} = \exp\left(-\left(\frac{b_u}{b_{ue}}\right)^2\right)$$

This gives $\frac{b_{ue}}{b_u} = 1.20$ and by virtue of equation (B3), $\alpha_{Eb_u} = 1.2\alpha_{Eb_{ue}}$.

For a jet,

$$\alpha_{Eb_{ue}} = 0.054 - 0.057 \text{ (see Turner 1986 and Hussain } et\ al.\text{ 1994)}.$$

This gives $\alpha_{Eb_u} = 0.065 - 0.068$.

For a plume,

$$\alpha_{Eb_{ue}} = 0.084 \text{ (see Turner 1986)}.$$

This gives $\alpha_{Eb_u} = 0.1$.